\documentclass{article}

\font\scap=cmcsc10 \hfuzz=5cm
\font\scap=cmcsc10

\usepackage{graphicx}

\newcount\eqnumber
\def\neweq{{\rm{(\the\eqnumber)}}\global\advance\eqnumber by 1}
\def\eqdef#1{\eqno\xdef#1{\the\eqnumber}\neweq}
\def\newaeq{{\rm{(\the\eqnumber a)}}\global\advance\eqnumber by 1}
\def\eqdaf#1{\eqno\xdef#1{\the\eqnumber}\newaeq}
\def\eqdisp#1{\xdef#1{\the\eqnumber}\neweq}
\def\eqdasp#1{\xdef#1{\the\eqnumber}\newaeq}

\newcount\refnumber
\def\newref{{\the\refnumber}\global\advance\refnumber by 1}
\def\refdef#1{{\xdef#1{\the\refnumber}}\newref}

\def\miso{\raise.5ex\hbox{$\scriptstyle 1$}\hbox{\kern-.12em$\scriptstyle /$}\kern-.12em\lower.5ex\hbox{$\scriptstyle 2$}}

\begin{document}

\centerline{\bf Restoring discrete Painlev\'e equations from an E$_8^{(1)}$-associated one}
\bigskip
\bigskip{\scap B. Grammaticos} and {\scap A. Ramani}
\quad{\sl IMNC,  CNRS, Universit\'e Paris-Diderot, Universit\'e Paris-Sud,\\ Universit\'e Paris-Saclay, 91405 Orsay, France}

\bigskip{\scap R. Willox}\quad
{\sl Graduate School of Mathematical Sciences, the University of Tokyo, 3-8-1 Komaba, Meguro-ku, 153-8914 Tokyo, Japan }
\medskip

{\scap Abstract}

\smallskip
We present a systematic method for the construction of discrete Painlev\'e equations. The method, dubbed `restoration', allows one to obtain all discrete Painlev\'e equations that share a common autonomous limit, up to homographic transformations, starting from any one of those limits. As the restoration process crucially depends on the classification of canonical forms for the mappings in the QRT family, it can in principle only be applied to mappings that belong to that family. However, as we show in this paper, it is still possible to obtain the results of the restoration even when the initial mapping is not of QRT type (at least for the system at hand, but we believe our approach to be of much wider applicability). For the equations derived in this paper we also show how, starting from a form where the independent variable advances one step at a time, one can obtain versions corresponding to multistep evolutions.

\bigskip

\bigskip
1. {\scap Introduction}

\medskip
Discrete Painlev\'e equations [\refdef\dps] are particularly rich integrable systems, to the point that calling them `Painlev\'e equations' may even sound reductive. There are just 6 continuous Painlev\'e equations while there exist infinitely many [\refdef\infin] discrete ones. The continuous Painlev\'e equations may involve up to four parameters but the discrete ones can have up to 7. So, where did the relation to Painlev\'e equations originate from? This goes back to the discovery of an integrable non-autonomous recursion relation by Brezin and Kazakov [\refdef\brezov]. Computing a partition function in a model of 2D quantum gravity they obtained the equation
$$x_{n+1}+x_n+x_{n-1}={z_n\over x_n}+1,\eqdef\zmed$$
where $z_n=\alpha n+\beta$, and found that for $z_n=-(3+2\epsilon^4t)/36$ and $x=(1-2\epsilon^2u)/6$, at the limit $\epsilon\to0$, this equation is nothing but the Painlev\'e I equation $u''=6u^2+t$. Thus (\zmed) was dubbed the discrete analogue of P$_{\rm I}$.

This naming convention was adopted in all subsequent studies, including those by the present authors: discrete equations were attributed a name based on the continuous  Painlev\'e equation that arose from them at the continuum limit. This was an unfortunate choice since, depending on the freedom one allows in a given discrete system one can in fact obtain different continuum limits. This is true for instance in the case of the Brezin-Kazakov system. More than a quarter-century before the latter authors, Shohat [\refdef\shohat] had already shown that the full expression of  the recursion relation (\zmed) was one with $z_n=\alpha n+\beta+\gamma(-1)^n$, the continuum limit of which  is Painlev\'e II. In fact, the presence of the $(-1)^n$ term is not innocuous. Discarding it on the na\"{\i}ve argument that ``$(-1)^n$ does not have a continuum limit'' is too simplistic. By now it is clear that the way to treat this term is by considering two copies of the equation, for even and for odd indices, and to take the continuum limit of the system of two first degree equations in two variables that ensues, only eliminating the extra continuous variable at the end of the process.

Of course the continuous and discrete Painlev\'e equations share many properties. {Both have solutions in terms of special functions for special values of their parameters, Schlesinger transformations can be constructed for those equations that possess at least one parameter and Miura transformations relating the various equations abound} [\refdef\pondy]. The solutions of all the discrete Painlev\'e equations possess the singularity confinement property [\refdef\sincon], which is the discrete analogue of the Painlev\'e property characterising the continuous Painlev\'e equations. A most important feature of the latter is that they are organised in a degeneration cascade [\refdef\capel], {from P$_{\rm VI}$ down to P$_{\rm I}$}. Quite expectedly the discrete Painlev\'e equations also follow a degeneration cascade --by the coalescence of parameters-- and this cascade provides the tool for the classification of these systems. As shown by Sakai [\refdef\sakai], the discrete Painlev\'e equations can be associated to affine Weyl groups, the latter  forming a degeneration pattern starting from the group E$_8^{(1)}$. {This pattern is given below (in its correct form, as will be explained further on) showing all possible degenerations, starting from} the elliptic discrete Painlev\'e equation all the way to the zero-parameter ones, explicitly including degenerations to multiplicative- and additive-type equations. The upper indices $e,q,d,c$ appearing in the names of the groups refer to the type of equations encountered in each of them, namely elliptic, multiplicative, difference and equations which are contiguity relations of continuous Painlev\'e equations.

\vskip.3cm
\centerline{{\includegraphics[width=16cm,keepaspectratio]{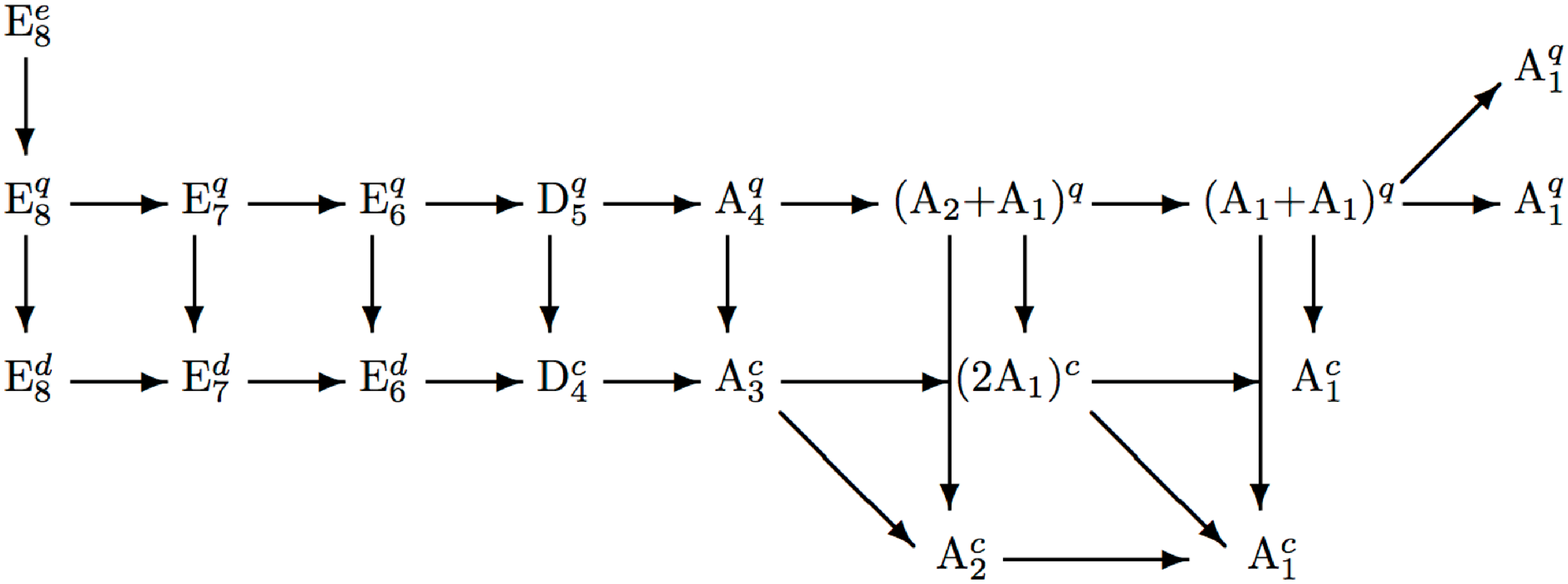}}}

After learning of Sakai's theory, we revised our original definition of the discrete Painlev\'e equations, {no longer referring to their continuum counterparts. Instead, according to} this new definition, a discrete Painlev\'e  equation is a mapping obtained by the periodic repetition of a non-closed pattern on a lattice associated to one of the affine Weyl groups (or subgroups thereof) belonging to the above degeneration cascade of E$_8^{(1)}$. An immediate consequence of this new definition is that the number of discrete Painlev\'e equations is infinite  [\infin], since any pattern, compatible with the above definition, would lead to a different discrete Painlev\'e equation. {Another consequence is that it also allows for a quite general discussion of the integrability of discrete Painlev\'e equations in terms of their degree growth. Takenawa [\refdef\take] has shown that all discrete Painlev\'e equations given in Sakai's paper [\sakai] have quadratic degree growth. Recently, it was shown by Mase [\refdef\mase] that any non-autonomous second order mapping with unbounded degree growth, that has the singularity confinement property and zero algebraic entropy, necessarily belongs to the class of mappings described in Sakai's classification (and that such a mapping can only have quadratic degree growth).}

Note that the degenerations to multiplicative and additive-type equations presented in the diagram above are not given in two similar diagrams in Sakai's seminal paper [\sakai]. In fact, although the diagrams in Sakai's paper are often interpreted as representing degeneration patterns for discrete Painlev\'e equations, this is actually not quite true. They represent all (mathematically) possible inclusion relations for  root subsystems of E$_8^{(1)}$, or for the orthogonal complements  thereof for special types of surfaces, which explains why they also contain arrows to cases that do not correspond to actual equations. 
When Sakai first presented his diagram we immediately noticed that {it contained} an arrow linking one of the A$_1^q$ groups to the lower of the A$_1^c$. However, if the diagram is interpreted as a degeneration pattern for actual equations such an arrow cannot exist. 
As a consequence the possible degeneration from $({\rm A}_1+{\rm A}_1)^q$ to the lower of the A$_1^c$ must be added (since in Sakai's diagram one could go from $({\rm A}_1+{\rm A}_1)^q$ to A$_1^c$ through A$_1^q$ and there was no need for such an arrow). Two of the present authors already presented the correct degeneration pattern, based on Sakai's diagram, in 2003 in the paper [\refdef\kundu]. Unfortunately this has gone essentially unnoticed and several authors have, over the years, been reproducing the diagram of Sakai in its incorrect interpretation. Note that our result has been corroborated by Rains [\refdef\rains] who, in 2016,  re-derived the degeneration pattern of the initial values spaces for discrete Painlev\'e equations, over fields {with characteristic zero}, which is the only case of interest to us (cf. Figure 1 in ref. [\rains]). 

Constructing new discrete Painlev\'e equations by the coalescence of parameters, starting from equations associated to the E$_8^{(1)}$ group as in the above degeneration procedure, is however not the only possibility. As we showed in [\refdef\resto] another process does exist. We call this the restoration method. In the next section we shall show how this method works and the remainder of the article will be devoted to a particularly rich application of the method to an asymmetric (in the QRT sense [\refdef\qrt]) equation that we first derived in [\refdef\deight].

\bigskip
2. {\scap The restoration method}

\medskip
The restoration method comprises two phases: a {\sl deconstruction} and a {\sl restoration} phase. We start with a discrete Painlev\'e equation, associated to some affine Weyl group as per the Sakai classification, and consider its autonomous limit. In most cases, as in the system we shall work with in this paper, our initial equation is related to the group E$_8^{(1)}$ but this is not mandatory: any discrete Painlev\'e equation could serve as an initial one. Next,  we neglect both the secular and periodic dependence on the independent variable in the coefficients of the equation, obtaining thus an autonomous mapping which, most often but not always (as we shall see below) belongs to the QRT family. Usually, at this stage, we then introduce an  obvious homographic transformation so as to have a {\sl remnant} equation that is as simple as possible.
The latter can be deautonomised but its deautonomisation most often leads to a discrete Painlev\'e equation associated to an affine Weyl group which, in the degeneration cascade of E$_8^{(1)}$, lies lower than that of the initial equation. In case the remnant mapping is a QRT mapping, we can obtain its invariant in one of the canonical forms we classified in [\refdef\auton, \refdef\canon]. This then ends the deconstruction phase, which provides us with an autonomous remnant equation and its invariant. The restoration phase starts with the introduction of a homographic transformation. In fact, we must first investigate which are the possible canonical forms the invariant can assume (other than the initial one) by considering different homographic transformations. Once a canonical form has been obtained, it is easy to write the corresponding QRT mapping, which can then be deautonomised to a discrete Painlev\'e equation by means of the singularity confinement integrability criterion.
Clearly, the equations thus obtained from the remnant equation will correspond to various Weyl groups in the Sakai classification. These {\sl restored} equations can be as simple as an A$_1^{(1)}$-associated one or as complicated as an elliptic Painlev\'e equation related to E$_8^{(1)}$.

This paper is devoted to the equations one can obtain, through the restoration process, starting from the asymmetric, additive E$_8^{(1)}$-associated, equation
$${x_{n+1}-(z_{n}+k_n)^2\over x_{n+1}-(z_{n}-k_n)^2}{x_{n}-(\zeta_n+k_n)^2\over x_{n}-(\zeta_n-k_n)^2}{y_{n}-(z_{n}+\zeta_n-k_n)^2\over y_{n}-(z_{n}+\zeta_n+k_n)^2}=1,\eqdaf\ddyo$$
$${y_{n}-(\zeta_{n-1}+\kappa_n)^2\over y_{n}-(\zeta_{n-1}-\kappa_n)^2}{y_{n-1}-(z_{n}+\kappa_n)^2\over y_{n-1}-(z_{n}-\kappa_n)^2}{x_n-(\zeta_{n-1}+z_n-\kappa_n)^2\over x_{n}-(\zeta_{n-1}+z_n+\kappa_n)^2}=1,\eqno(\ddyo b)$$
where $z_n=\eta_{n-1}-\gamma$, $\zeta_n=\eta_{n+1}+\gamma$, $k_n=\eta_{n+1}+\eta_{n}+\eta_{n-1}-\phi_2(n)$, $\kappa_n=\gamma+\phi_2(n)$ with $\eta_n=\alpha n+\beta+\phi_3(n)+\phi_4(n)$. Here $\phi_m$ is a periodic function, with period $m$, given by 
 $$ \phi_m(n)=\sum_{l=1}^{m-1} \delta_l^{(m)} \exp\left({2i\pi ln\over m}\right).\eqdef\dtri$$
Note that the summation starts at 1 instead of 0 and thus $\phi_m$ introduces $m-1$ parameters. Anticipating on what we shall encounter later in this paper, we introduce also the periodic function $\chi_{2m}$ with period $2m$, involving only $m$ parameters, which can be expressed in terms of roots of unity as
$$ \chi_{2m}(n)=\sum_{\ell=1}^{m} \eta_{\ell}^{(m)} \exp\left({i\pi(2\ell-1)n\over m}\right).\eqdef\dtra$$
Equation (\ddyo) was first derived in [\deight], and referred to in [\refdef\multi] as case IX. 

Neglecting the periodic dependence and keeping only the secular one we can rewrite (\ddyo) as
$${x_{n+1}-(4\eta_n-\alpha- \gamma)^2\over x_{n+1}-(2\eta_n+\alpha+ \gamma)^2}{x_{n}-(4\eta_n+\alpha+ \gamma)^2\over x_{n}-(2\eta_n-\alpha- \gamma)^2}{y_{n}-\eta_n^2\over y_{n}-25\eta_n^2}=1,\eqdaf\dtes$$
$${y_{n}-(\eta_n+2 \gamma)^2\over y_{n}-\eta_n^2}{y_{n-1}-(\eta_n-\alpha)^2\over y_{n-1}-(\eta_n-\alpha-2 \gamma)^2}{x_{n}-(2\eta_n-\alpha- \gamma)^2\over x_{n}-(2\eta_n-\alpha+ \gamma)^2}=1.\eqno(\dtes b)$$
Equation (\dtes) has four distinct singularity patterns (and the same is true for (\ddyo)) 
of lengths 1,1,4 and 10 (where we are counting the number of steps of the pattern). They are: $\{x_n=(2\eta_n-\alpha- \gamma)^2,y_n=\eta_n^2\}$, $\{y_n=\eta_n^2,x_{n+1}=(2\eta_{n+1}-\alpha+ \gamma)^2\}$, $\{x_{n-1}=(2\eta_{n-1}-\alpha+ \gamma)^2,y_{n-1}=(\eta_{n-1}+2 \gamma)^2,x_n=( \gamma-1)^2,y_n=(\eta_{n+1}-2 \gamma)^2,x_{n+1}=(2\eta_{n+1}-\alpha- \gamma)^2\}$ and $\{y_{n-2}=25\eta_{n-2}^2, x_{n-1}=(4\eta_{n-2}-\alpha- \gamma)^2,y_{n-1}=(3\eta_{n-2}-\alpha)^2,x_{n}=(2\eta_{n-2}-3\alpha- \gamma)^2,y_{n}=(\eta_{n-2}-4\alpha)^2,x_{n+1}=( \gamma+7\alpha)^2,y_{n+1}=(\eta_{n+3}+4\alpha)^2,x_{n+2}=(2\eta_{n+3}+3\alpha+ \gamma)^2,y_{n+2}=(3\eta_{n+3}+\alpha)^2,x_{n+3}=(4\eta_{n+3}+\alpha+ \gamma)^2,y_{n+3}=25\eta_{n+3}^2\}$. 

Neglecting the secular dependence by taking $\alpha=0$ and scaling $\beta$ to 1 we have
$${x_{n+1}-(4-\gamma)^2\over x_{n+1}-(2+ \gamma)^2}{x_{n}-(4+ \gamma)^2\over x_{n}-(2- \gamma)^2}{y_{n}-1\over y_{n}-25}=1,\eqdaf\dpen$$
$${y_{n}-(1+2 \gamma)^2\over y_{n}-1}{y_{n-1}-1\over y_{n-1}-(1-2 \gamma)^2}{x_{n}-(2- \gamma)^2\over x_{n}-(2+ \gamma)^2}=1.\eqno(\dpen b)$$
While (\dpen) is indeed autonomous it is not quite suitable for the restoration process since, due to the presence of $\gamma$, it is not of QRT type. We have in fact {\sl two} `invariants': $K_1(x_n,y_n)$ given by
$K_1(x_n,y_n)={\cal K}(x_n,y_n;\gamma)$  where
$$\displaylines{\Big((x_n-y_n)^2-2(\gamma-1)^2(x_n+y_n)+(\gamma-1)^4\Big){\cal K}(x_n,y_n;\gamma)=\hfill\cr\hfill x_n^2y_n^2(\gamma^2 - 2\gamma + 1) + 2x_n^2y_n( - 9\gamma^2 + 10\gamma - 1) + 9x_n^2(\gamma^2 - 2\gamma - 3) + 2x_ny_n^2( - \gamma^4 + \gamma^2 + 10\gamma - 10) \hfill\cr\hfill+ 4x_ny_n(9\gamma^4 - 8\gamma^3 + 35\gamma^2 - 42\gamma + 24) + 18x_n( - \gamma^4 - 3\gamma^2 + 2\gamma + 2) + y_n^2(\gamma^6 + 2\gamma^5 - 3\gamma^4 - 20\gamma^3 - 52\gamma^2 + 36) \hfill\cr\hfill+ 6y_n( - 3\gamma^6 + 2\gamma^5 + 25\gamma^4 - 20\gamma^3 - 16\gamma^2 + 24\gamma - 12) + 9(\gamma^6 + 2\gamma^5 - 23\gamma^4 + 28\gamma^3 + 4\gamma^2 - 16\gamma + 4),\quad\eqdisp\arfuno\cr}$$
and  $K_2(x_n,y_{n-1})={\cal K}(x_n,y_{n-1},-\gamma)$.
Both are bi-quadratic (as in the QRT case) but in this case the conservation law is {\sl not} the usual QRT one but rather
$$K_2(x_n,y_{n-1}) =K_1(x_n,y_n) =K_2(x_{n+1},y_n),\eqdef\arftres $$
resulting to (\dpen) (and two spurious equations which need not concern us here: they go over to the trivial equations $x_n=x_{n+1}$, $y_{n-1} =y_n$ at the $\gamma\to0$ limit). 

Since (\dpen) is not of QRT type we simplify it further by taking $\gamma\to0$. Computing the appropriate limit we find
$${x_{n+1}-16\over x_{n+1}-4}{x_{n}-16\over x_{n}-4}{y_{n}-1\over y_{n}-25}=1,\eqdaf\dhex$$
$${1\over y_{n}-1}+{1\over y_{n-1}-1}-{2\over x_{n}-4}=0.\eqno(\dhex b)$$
Next, we introduce the homographies
$$X_n={1\over4}{x_{n}-16\over x_{n}-4}, \quad Y_n={1\over16}{y_{n}-25\over y_{n}-1},\eqdef\dhep$$
which allows us to write (\dhex) as
$$X_{n+1}X_n=Y_n,\eqdaf\doct$$
$$Y_n+Y_{n-1}=X_n-A.\eqno(\doct b)$$
This is the remnant equation  we shall use for the reconstruction procedure. 
Note that, here, $A$ has the precise numerical value 1/8, but in what follows we shall consider it to be a free parameter.

A remark is in order at this point. Case IX of [\multi] is not the only E$_8^{(1)}$-associated asymmetric trihomographic equation with singularity patterns of length 1,1,4 and 10.  In the derivation presented in [\deight] we obtained also an equation, dubbed case VIII in [\multi], where $z_n=\eta_{n-1}+\psi(n-1)$, $\zeta_n=\eta_{n+1}-\psi(n)$, $k_n=\eta_{n+1}+\eta_{n}+\eta_{n-1}+\psi(n)-\psi(n-1)$, $\kappa_n=\psi(n+1)+\psi(n)$ where $\eta_n=\alpha n+\beta+\phi_3(n)+\phi_4(n)$ and $\psi(n)=\delta_1(-j)^n+\delta_2(-j)^{2n}$, ($j^3=1$) in (\ddyo). At the autonomous limit, when only $\beta$ survives in the coefficients, the remnant equation obtained from case VIII is precisely the one obtained from the autonomous limit of case IX, i.e. equation (\doct). But the relation between VIII and IX goes even deeper than that. In fact, if we neglect the $\psi(n)$ and $\phi_2(n)$ term in VIII and IX respectively and put $\gamma=0$ in the latter we obtain the very same discrete Painlev\'e equation, since the expression of $\eta_n$ is the same for the two systems.

 The mapping (\doct) being of QRT type, we can easily obtain its invariant: 
$$K={X_nY_n(X_n-Y_n)-(A+1)X_nY_n+Y_n^2+AX_n^2+AY_n\over X_n}.\eqdef\denn$$
Note that, given the form of (\doct) we could have eliminated either of the two variables and obtain a symmetric three-point mapping. However in what follows we opt to work with the asymmetric form (\doct).
 
Equation (\doct) has a confined singularity pattern. Starting form a finite $X_0$ and $Y_0=0$ we find the following pattern $\{0,0,-A,\infty,\infty,1,\infty,\infty,-A,0,0\}$. Moreover two cyclic patterns do exist: $\{Y_0,0,f,\infty,\infty,1,\infty,\infty\}$ ($f$ being a finite value depending on $Y_0$) and $\{X_0,\infty,\infty,\infty\}$. Deautonomising (\doct) is straightforward. We assume that $A$ is a function of $n$ and require that the confined singularity pattern remain as such. We thus find the constraint 
$$A_{n+4}-A_{n+3}-A_{n+1}+A_n=0,\eqdef\ddek$$ 
the solution of which is $A_n=\alpha n+\beta+\phi_3(n)$. This non-autonomous form of equation (\doct) is well-known: it was first derived in [\refdef\dress]. The number of genuine degrees of freedom, i.e. those that can be modified through the application of Schlesinger transformations,
 is 3, namely $\beta$ and the two parameters introduced by $\phi_3$, ($\alpha$  is essentially the step of the Schlesinger transformation and does not count).
The geometry of its transformations is described by the affine Weyl group A$_3^{(1)}$. Eliminating the variable $Y$ we obtain for $X$ the equation
$$X_{n+1}+X_{n-1}=1+{A_n\over X_n},\eqdef\bena$$
which is the well-known contiguity relation for the solutions of the continuous P$_{\rm V}$ equation [\refdef\cfive]. Similarly, eliminating $X$ we find an equation for $Y$ which we can write, after the translation $Y_n\to Y_n+Z_n$ with $A_n=Z_n+Z_{n-1}$, as
$$(Y_{n+1}+Y_n)(Y_n+Y_{n-1})=Y_n+Z_n.\eqdef\bdyo$$
an equation first derived in [\refdef\japan].
The system (\doct) can thus be considered as the Miura transformation between (\bena) and (\bdyo). A double-step evolution can be obtained from (\bena). We translate $X_n\to X_n+\miso$ and find readily the equation
$${A_{n+1}\over X_{n+2}+X_n}+{A_{n-1}\over X_n+X_{n-2}}=1+{A_n\over X_n+\miso},\eqdef\btri$$
an equation already obtained in [\refdef\congr]. A triple-step evolution, starting from (\doct), is also possible. We find the system
$$X_{n-2}X_{n+1}={(Y_{n-1}+A_n)(Y_{n-1}+A_{n-1})\over Y_{n-1}},\eqdaf\btes$$
$$Y_{n+2}+Y_{n-1}=-X_{n+1}-{A_{n-1}\over 1+X_{n+1}}-A_{n}-A_{n+2}+A_{n+1}.\eqno(\btes b)$$
It can be shown that, after a few elementary transformations, this equation coincides with one we derived in collaboration with H. Sakai, equation (4) in [\refdef\ohta].

\bigskip
3. {\scap Restoring discrete Painlev\'e equations}

\medskip
Many more discrete Painlev\'e equations can be found however if we apply appropriate homographies to the remnant equation.
Starting from (\doct) and its invariant (\denn), and adding a constant $\mu$ to the latter to allow for an arbitrary value of the conserved quantity, we introduce a homographic transformation of the dependent variables in the form
$$X_n={ax_n+b\over cx_n+d},\eqdaf\vena$$
$$Y_n={py_n+q\over ry_n+s},\eqno(\vena b)$$
and try to match the resulting form of the invariant to one of the canonical QRT forms. 

Before applying the restoration process for the obtention of new equations we remark that, choosing the homography (\dhep), with $A=1/8$, we obviously can go back to our starting point, namely the additive, E$_8^{(1)}$-associated discrete Painlev\'e equation (\ddyo). 
Eliminating $y_n$ from the system (\ddyo) we obtain a `symmetric', in the QRT sense, trihomographic equation for $x_n$, which is precisely equation V given in our paper [\multi]. Eliminating $x_n$ we obtain for $y_n$ an equation which, when deautonomised, is precisely equation 4.5.1 of our paper [\refdef\addit]. Since the symmetric equation for $x$ is trihomographic we can easily construct the equation corresponding to a double-step evolution. We shall not go here into these details. Suffice it to say that when deautonomised this equation is the case 5.2.1 we identified in [\addit]. 

Since the remnant equation for case VIII is the same as that for case IX it is clear that we can, by restoration, obtain also the E$_8^{(1)}$-associated equation corresponding to the former. As this equation is trihomographic, we can eliminate either of the variables $x$ or $y$ and obtain an equation for a single variable. Eliminating $x_n$ we obtain for $y_n$ an equation which coincides with the one obtained from case IX i.e. equation 4.5.1 of our paper [\addit]. However, when we eliminate $y_n$ we do not obtain case V but rather equation 4.4.4 of [\addit] (which is not trihomographic, making a double-step evolution impossible). We should point out here that equation 4.4.4 exists only if the quantity $\psi(n)$, introduced in the previous section in relation to VIII, does not vanish. Were $\psi(n)$ to vanish, the resulting equation would collapse to case V with $k_n$ identically zero.

The first restoration we shall consider is towards the multiplicative E$_8^{(1)}$-associated equation. Note that, while we are working with the $\gamma=0$ case, what we obtain, upon restoration, is the full multiplicative equation, i.e. the multiplicative analogue of case V and case II (which are essentially identical when fully non-autonomous, as we shall explain in the next section). Parametrising $A$ as
$$A={z^4+1/z^4\over (z^2+1/z^2)^4}\eqdef\vdyo$$
and introducing the transformation
$$X_n={1\over(z^2+1/z^2)^4}{x_n+z^{10}+1/z^{10}\over x_n+z^2+1/z^2},\eqdaf\vtri$$
$$Y_n={1\over(z^2+1/z^2)^2}{y_n+z^8+1/z^8\over y_n+z^4+1/z^4},\eqno(\vtri b)$$
we obtain the invariant
$$K={(x_n+z^2+1/z^2)^2(y_n+z^4+1/z^4)(y_n+z^8+1/z^8)\over x_n^2-(z^2+1/z^2)x_ny_n+y_n^2+(z^2-1/z^2)^2},\eqdef\vtes$$
using $\mu=(z^8+z^4+1+1/z^4+1/z^8)/(z^2+1/z^2)^8$ which leads to a mapping of multiplicative type. The deautonomisation of this mapping (and the extension to the elliptic case) can be directly obtained, based on that of (\ddyo), following the method described in [\multi]. 

While it is not possible to find equations associated to the group E$_7^{(1)}$ one can obtain equations related to E$_6^{(1)}$.  Taking, for instance, the homography
$$X_n={x_n-A\over x_n+1},\eqdaf\vpen$$
$$Y_n={-Ay_n+1\over y_n+1},\eqno(\vpen b)$$
we obtain the invariant
$$K={(y_n+1)^2(x_n+1)(x_n-A)\over x_ny_n-1},\eqdef\vhex$$
for $\mu=A$. 
The resulting mapping has the form
$$(y_{n}x_n-1)(y_nx_{n+1}-1)=-Ay_n^2-(A-1)y_n+1,\eqdaf\vhep$$
$$(y_{n-1}x_n-1)(y_nx_{n}-1)=x_n^2+2x_n+1.\eqno(\vhep b)$$
This is a special, somewhat parameter-poor, form of an equation studied in detail in [\refdef\qasym]. In order to deautonomise it we start by introducing more parameters and rewrite (\vhep) as
$$(y_{n}x_n-1)(y_nx_{n+1}-1)=(1-a y_n)(1-b y_n),\eqdaf\voct$$
$$(y_{n-1}x_n-1)(y_nx_{n}-1)=(1-cx_n)(1-dx_n).\eqno(\voct b)$$
Using singularity confinement we find that the non-autonomous form of (\voct) is
$$(y_{n}x_n-1)(y_nx_{n+1}-1)=(1-a_n y_n)(1-b_n y_n),\eqdaf\venn$$
$$(y_{n-1}x_n-1)(y_nx_{n}-1)=(1-x_n/b_n)(1-x_n/b_{n-1}),\eqno(\venn b)$$
where $\log a_n=\alpha n+\beta+\phi_4(n)$, $\log b_n=\gamma n+\delta+\phi_2(n)$. Using the gauge freedom in the equations we can show that only the difference $\beta-\delta$ is meaningful, so the coefficients in the equation only contain 6 genuine parameters. Indeed, although $\gamma$ enters through the product $\gamma n$ it is a genuine degree of freedom: it can be modified through the application of Schlesinger transformations, and this is in contradistinction to $\alpha$, which also enters through $\alpha n$, but is essentially the step of the Schlesinger transformation.

It is now interesting to look at the equations one obtains for $x$ and $y$ alone. Eliminating the variable $x$ we find 
$$(b_{n}b_{n-1}y_{n-1}y_n-1)(b_{n}b_{n+1}y_ny_{n+1}-1)={(1-b_ny_n)(1-b_{n+1}y_n)(1-b_{n-1}y_n)\over 1-a_ny_n},\eqdef\vdek$$
and, introducing  the gauge $b_ny_n\to y_n$, we rewrite (\vdek) as
$$(y_{n-1}y_n-1)(y_ny_{n+1}-1)={(1-y_n)(1-b_{n+1}y_n/b_n)(1-b_{n-1}y_n/b_n)\over 1-a_ny_n/b_n},\eqdef\vend$$
an equation already obtained in [\qasym], equation A1cix.
Similarly eliminating $y$ we find the equation
$$(x_{n-1}x_n-a_{n-1}b_{n-1})(x_nx_{n+1}-a_{n}b_{n})=b_{n}b_{n-1}(x_n-a_n)(x_n-a_{n-1}),\eqdef\vdod$$
and a gauge could have been chosen in order to bring it to canonical form but this would require the introduction of terms of the type $n\phi_2(n)$ in the coefficients (which, as a general rule, we try to avoid). Equation (\voct) is thus the Miura relation linking (\vdek) and (\vend). And, of course, it is also a discrete Painlev\'e equation in its own right. 

Given that in (\vdod) the constant term cancels out and after division the equation is linear in $x_n$ it is straightforward to obtain the double-step evolution.  We find readily
$$(x_{n-2}x_n-b_{n-2}b_{n-1})(x_nx_{n+2}-b_{n}b_{n+1})={(a_{n+1}x_n-a_nb_n)(a_{n-2}x_n-a_{n-1}b_{n-1})(x_n-b_n)(x_n-b_{n-1})\over(x_n-a_n)(x_n-a_{n-1})}.\eqdef\ttri$$
This is the generic E$_6^{(1)}$-associated equation [\qasym] and can be cast into canonical form by means of a rather complicated gauge transformation.
An equation corresponding to a triple-step evolution can also be obtained but as in the previous case it can only be expressed in asymmetric form. We find
the system
$$\left({x_{n-2}-a_{n-2}b_{n-2}y_{n-1}\over x_{n-2}-b_{n-2}b_{n-1}y_{n-1}}\right)\left({x_{n+1}-a_{n}b_{n}y_{n-1}\over x_{n+1}-b_{n}b_{n-1}y_{n-1}}\right)={1-a_{n-1}y_{n-1}\over 1-b_{n-1}y_{n-1}},\eqdaf\ttes$$
$$\left({x_{n+1}-a_{n}b_{n}y_{n-1}\over x_{n+1}-b_{n}b_{n-1}y_{n-1}}\right)\left({x_{n+1}-a_{n+1}b_{n+1}y_{n+2}\over x_{n+1}-b_{n+1}b_{n+2}y_{n+2}}\right)={(x_{n+1}-a_{n+1})(x_{n+1}-a_{n})\over (x_{n+1}-b_{n+1})(x_{n+1}-b_n)}.\eqno(\ttes b)$$
 This is a discrete Painlev\'e equation, in a slightly non canonical form of systems belonging to the family known as VI$'$ in our classification [\canon] of canonical forms. Note that although the generic equations of this family are associated with the group E$_7^{(1)}$, here we are nevertheless in the presence of an equation related to E$_6^{(1)}$.

One could of course also expect to find additive equations, also related to E$_6^{(1)}$, for special values of the parameter $A$. We find indeed that when $A=-1$, introducing the homographic transformation
$$Y_n={y_n-q\over y_n-1},\eqdaf\tpen$$
$$X_n={x_n+q\over x_n+1},\eqno(\tpen b)$$
we obtain the invariant
$$K={(y_n-1)^2(x_n+1)(x_n+q)\over y_n+x_n},\eqdef\thex$$
for $\mu=-1$. The corresponding mapping is
$$(y_{n}+x_n)(y_n+x_{n+1})=y_n^2-(1+q)y_n+q,\eqdaf\thep$$
$$(y_{n-1}+x_n)(y_n+x_{n})=x_n^2+2x_n+1,\eqno(\thep b)$$
which was also studied in [\refdef\dasym]. Again, in order to deautonomise it, we enrich it by introducing more parameters and write
$$(y_{n}+x_n)(y_n+x_{n+1})=(y_n-a)(y_n-b),\eqdaf\toct$$
$$(y_{n-1}+x_n)(y_n+x_{n})=(x_n+c)(x_n+d).\eqno(\toct b)$$
By application of singularity confinement we find
$$(y_{n}+x_n)(y_n+x_{n+1})=(y_n-a_n)(y_n-b_n),\eqdaf\tenn$$
$$(y_{n-1}+x_n)(y_n+x_{n})=(x_n+b_n)(x_n+b_{n-1}).\eqno(\tenn b)$$
again with $a_n=\alpha n+\beta+\phi_4(n)$ and $b_n=\gamma n+\delta+\phi_2(n)$.
Eliminating $x_n$ we find, after translating $y_n$ by $-b_n$,
$$(y_{n-1}+y_n)(y_n+y_{n+1})={y_n(y_n-(b_{n+1}-b_n))(y_n-(b_{n-1}-b_n))\over y_n-(a_n-b_n)},\eqdef\tdek$$
already obtained in [\dasym], equation (42) case A4d in that reference.
Similarly for $x$ we find
$$(x_{n-1}+x_n+a_{n-1}+b_{n-1})(x_n+x_{n+1}+a_{n}+b_{n})=(x_n+a_n)(x_n+a_{n-1}),\eqdef\qena$$
and again, the gauge that would allows us to bring (\qena) to canonical form involves $n\phi_2(n)$, which, as already explained, is something we try to avoid. 

As in the case of the multiplicative equations, system (\tenn) is a Miura relating (\tdek) and (\qena). Double- and triple-step evolutions can also be obtained just as in the multiplicative case.  We find for instance
$$(x_{n-2}+x_n+b_{n-2}+b_{n-1})(x_n x_{n+2}+b_{n}+b_{n+1})\!=\!{(x_n-a_{n+1}+a_n+b_n)(x_n-a_{n-2}+a_{n-1}+b_{n-1})(x_n+b_n)(x_n+b_{n-1})\over(x_n+a_n)(x_n+a_{n-1})}\eqdef\rafone$$
for the double-step evolution which is the generic additive E$_6^{(1)}$-associated equation. The triple step evolution is given as the system
$$\left({x_{n-2}-y_{n-1}+a_{n-2}+b_{n-2}\over x_{n-2}-y_{n-1}+b_{n-2}+b_{n-1}}\right)\left({x_{n+1}-y_{n-1}+a_{n}+b_{n}\over x_{n+1}-y_{n-1}+b_{n}+b_{n-1}}\right)={y_{n-1}-a_{n-1}\over y_{n-1}-b_{n-1}},\eqdaf\raftwo$$
$$\left({x_{n+1}-y_{n-1}+a_{n}+b_{n}\over x_{n+1}-y_{n-1}+b_{n}+b_{n-1}}\right)\left({x_{n+1}-y_{n+2}+a_{n+1}+b_{n+1}\over x_{n+1}-y_{n+2}+b_{n+1}+b_{n+2}}\right)={(x_{n+1}+a_{n+1})(x_{n+1}+a_{n})\over (x_{n+1}+b_{n+1})(x_{n+1}+b_n)}.\eqno(\raftwo b)$$
which by changing $y$ to $-y$ and introducing the appropriate gauge can be cast to a canonical form.

\bigskip
4. {\scap The non-zero-$\gamma$ case}

\medskip
As we explained in section 2, we cannot simply apply the restoration process described above to the autonomous mapping (\dpen) since the latter, due to the presence of $\gamma$, is not of QRT type. However something can still be done. Given that both equations are in trihomographic form we can eliminate either $x$ or $y$ and obtain a QRT-type mapping for a single variable. We start thus from
$${x_{n+1}-(4z-\gamma)^2\over x_{n+1}-(2z+ \gamma)^2}{x_{n}-(4z+ \gamma)^2\over x_{n}-(2z- \gamma)^2}{y_{n}-z^2\over y_{n}-25z^2}=1,\eqdaf\qdyo $$
$${y_{n}-(z+2 \gamma)^2\over y_{n}-z^2}{y_{n-1}-z^2\over y_{n-1}-(z-2 \gamma)^2}{x_{n}-(2z- \gamma)^2\over x_{n}-(2z+ \gamma)^2}=1,\eqno(\qdyo b)$$ 
and eliminate in turn $y$ and $x$. We find readily for $x$ the equation
$${(x_{n+1}-x_n-4z^2)(x_{n-1}-x_n-4z^2)-16z^2x_n\over x_{n+1}-2x_n+x_{n-1}-8z^2}=-{x_n+16z^2-\gamma^2\over2},\eqdef\qtri$$
which, when deautonomised, is precisely the equation we called II in [\multi]. And, of course, when we consider the double-step evolution for (\qtri) we find again an equation that can be deautonomised to the case 5.2.1 of [\addit] just as in section 3.

An important remark must be made here. When examining the $\gamma=0$ case in section 3, we remarked that the E$_8^{(1)}$-related equation obtained for $x_n$ alone was equation V of [\multi]. However the case V and the case II, referred to just above, are {\sl not} different. As shown in [\multi], for both cases II and V the $n$-dependence was obtained with $\eta_n=\alpha n+\beta+\phi_2(n)+\phi_3(n)$  and  $z_n+z_{n+1}=\eta_{n+1}$ but for case II we found $k_n=\gamma+\phi_4(n)$ while for case V we had $k_n=\chi_8(n)$. (Note that our convention for $\chi_m$ has changed between [\multi] and [\addit] and we are now using the one introduced in the latter, so what was $\chi_4$ in the former paper is now given as $\chi_8$).
Looking closely, these two $k_n$'s are just two different ways of presenting the same result. The $k_n=\chi_8(n)$ introduces four parameters which alternate sign every four steps. The $k_n=\gamma+\phi_4(n)$ introduces four parameters which keep repeating every four steps. However changing the sign of $k_n$ for some $n$ does not change anything: it corresponds to just taking the inverse of each side of the equation. So case V and case II are the same. The fact that the former was obtained for $\gamma=0$ and the latter for $\gamma\ne0$ does not have any deep meaning. When one deautonomises a mapping completely, one must recover the full complement of parameters. Thus, although in section 3 we started with $\gamma=0$, the deautonomisation led to $k_n=\chi_8(n)$ and thus to four parameters, as expected, for the equation to have its full complement of degrees of freedom. However, when taking the remnant through the autonomous form, the zero or nonzero value of $\gamma$ strongly affects the result: the remnant is QRT only if $\gamma=0$. 

The remnant system for $\gamma\neq0$ can be obtained in a straightforward way. We introduce a new variable $X$ through
$$X_n=\left({3\beta-2\gamma\over 3\beta-6\gamma}\right){x_n-(\beta+\gamma)^2\over x_n-(\beta-\gamma)^2},\eqdef\vdek$$
and rewrite (\qtri) as
$$X_{n+1}X_{n-1}=A{X_n-B\over X_n-1}\eqdef\tena$$
where $A$ and $B$ have precise expressions in terms of $\beta$ and $\gamma$ and are, in fact, related. However, relinquishing this relation between $A$ and $B$ one can go back to the variable $x_n$ and obtain a (full parameter) multiplicative E$_8^{(1)}$-associated equation, precisely the $q$-analogue of (\qtri) (and an elliptic equation can also be obtained along the same lines). 

The invariant of (\tena) is
$$K={(X_n-A)(X_{n-1}-A)(X_nX_{n-1}-X_n-X_{n-1}+B)\over X_nX_{n-1}}.\eqdef\tdyo$$
 Deautonomising (\tena) we find $\log A_n= \alpha n+\beta+\phi_2(n)+\phi_3(n)$ and $\log B_n= 2\alpha n+\gamma-\phi_3(n)$ leading to a discrete Painlev\'e equation associated to the group D$_5^{(1)}$. A double-step evolution for $X$ can be obtained almost by inspection of (\tena). We find readily
 $$\left({X_{n+2}X_n-A_{n+1}B_{n+1}\over X_{n+2}X_n-A_{n+1}}\right)\left({X_{n-2}X_n-A_{n-1}B_{n-1}\over X_{n-2}X_n-A_{n-1}}\right)=A_n{X_n-B_n\over X_n-1},\eqdef\ttri$$
which was already obtained in [\refdef\sequel] (equation 2.20, but written here in a different gauge).

Eliminating $x$ in (\qdyo) we obtain for $y$ he equation
$${(y_{n+1}-y_n-4\beta^2)(y_{n-1}-y_n-4\beta^2)-16\beta^2y_n\over y_{n+1}-2y_n+y_{n-1}-8\beta^2}=-{y_n^2+2y_n(9\beta^2-2\gamma^2)+5\beta^2(\beta^2-4\gamma^2)\over2y_n+4\beta^2-6\gamma^2}.\eqdef\pena$$
The deautonomisation of (\pena) to an additive discrete Painlev\'e equation was presented in [\addit], case 4.5.1. (Given the remark we made at the beginning of this section it is not astonishing that the deautonomisation of (\voct) is exactly the same as the one obtained for the $y$ additive E$_8^{(1)}$-associated equation in section 3). Introducing, admittedly with hindsight, the homographic transformation
$$Y_n=g{y_n(\gamma^2+3\beta^2)-3\beta^2(11\gamma^2+\beta^2)\over y_n-\beta^2}\eqdef\pdyo$$
and, choosing the appropriate value for $g$, we find from (\pena)
$$(Y_{n+1}Y_n-1)(Y_nY_{n-1}-1)=(1-CY_n)(1-Y_n/C)(1-Y_n/D)\eqdef\ptri$$
where $C$ and $D$ are expressed in terms of $\gamma$ and $\beta$. This is the remnant equation. Its invariant has the form
$$K={(1-DY_n)(1-DY_{n-1})(Y_n+Y_{n-1}-C-1/C)\over Y_nY_{n-1}-1}.\eqdef\ptes$$
Deautonomising (\ptri) we find that $C$ is constant and  $D_n=\alpha n+\beta+\phi_2(n)+\phi_3(n)$ (equation (62) in [\qasym]). This is an equation associated to D$_5^{(1)}$ just as the deautonomised (\tena).

What is particularly interesting at this point is that we can establish a Miura relation between equation (\tena) and equation (\ptri). 
$$X_n=D_n{CY_{n-1}-1\over C-Y_n},\eqdaf\ppen$$
$$Y_n={X_{n+1}(X_n-1)+D_nD_{n+1}\over X_nD_{n+1}},\eqno(\ppen b)$$
where $A,B$ are related to $C,D$ by $A_n=CD_{n+1}$, $B_n=D_nD_{n-1}$. Given that $X$ and $Y$ are related to the initial system (\qdyo) through simple homographic transformations, the Miura relation (\ppen) can be considered as D$_5^{(1)}$-related deautonomisation of (\qdyo). Thus working on the mappings for $x$ and $y$ separately we were able to obtain the result we could not obtain directly on the non-QRT mapping (\qdyo).
   
Starting from (\ppen) we can also obtain a triple-step evolution. We find thus the system
$$\displaylines{(Y_{n-1}-D_{n-1})(CD_nD_{n+1}X_{n-2}Y_{n-1}+X_{n-2}X_{n+1}-CD_{n}D_{n-1}D_{n-2}X_{n+1}-C^2D_nD_{n+1})\hfill\cr\hfill-CX_{n+1}(D_nD_{n-1}-1)(D_{n-1}D_{n-2}-1)=0,\quad\eqdasp\phex\cr}$$
$$CD_nD_{n+1}Y_{n-1}(X_{n+1}-1)+CY_{n+2}X_{n+1}(X_{n+1}-D_nD_{n+1})-X_{n+1}^2-(C^2+1)X_{n+1}+C^2D_nD_{n+1}=0.\eqno(\phex b)$$

Next we seek a restoration to equations related to E$_6^{(1)}$. We start from the remnant equation (\tena) and the associated invariant (\tdyo). Introducing the homography
$$X_n={Ax_n+A-B\over x_n+A-1}\eqdef\phep$$
we obtain the invariant
$$K={(Ax_n+A-B)( x_n+A-1)(Ax_{n-1}+A-B)( x_{n-1}+A-1)\over x_nx_{n-1}+B-1},\eqdef\poct$$
with $\mu=0$. Introducing $B=1-c^2$ and rescaling $x\to cx$ we find the equation
$$(x_{n+1}x_n-1)(x_nx_{n-1}-1)=(1-ax_n)(1-bx_n),\eqdef\penn$$
where $A=(a^2-1)/(ab-1)$ and $c=(b-a)/(ab-1)$.
Similarly for $Y$ we start from (\ptri) and (\ptes) and introduce
$$Y_n={(y_n-1)/D+C+1/C\over y_n+1}\eqdef\pdek$$
which leads to the invariant
$$K={(y_n+1)(y_{n-1}+1)(y_ny_{n-1}(1-D^2)-(y_n+y_{n-1})(D-C)(D-1/C)+(DC+D/C-1)^2-D^2)\over y_ny_{n-1}-1}\eqdef\sena$$
(again with $\mu=0$) from which we obtain the equation
$$(y_{n+1}y_n-1)(y_ny_{n-1}-1)={(y_n+1)(y_n-f)(y_n-1/f)\over 1-dy_n},\eqdef\sdyo$$
with $D=C(d-f)/(d+1)$ and $C^2=(d-1/f)/(d-f)$. Given the relations between $A,B,C,D$ we find readily $f=-b/a$ and $d=-1/(ab)$. We could now proceed to the  deautonomisation of  the equations (\penn) and (\sdyo). However this is not necessary, since a simple glance suffices to convince us that this would lead to the forms (\vend) and (\vdod) already obtained in the $\gamma=0$ case. Thus in the case of E$_6^{(1)}$-associated equations the results are exactly the same, whether $\gamma$ is 0 or not, the Miura relation in both cases being (\venn). However, what is essential is that in the $\gamma\ne0$ case we were able to implement the restoration procedure starting from the non-QRT mapping (\qdyo).

Obviously, additive equations related to E$_6^{(1)}$ can  also be constructed. Starting from the invariant (\tdyo) and introducing the transformation
$$X_n={Ax_n+A-2\over x_n-1},\eqdef\uena$$
we find the invariant
$$K={(x_n-1)(x_{n-1}-1)(Ax_n+A-2)(Ax_{n-1}+A-2)\over x_n+x_{n-1}},\eqdef\udyo$$
with $\mu=0$, provided the parameters $A,B$ obey the constraint $B=2A-A^2$. The resulting equation is
$$(x_{n+1}+x_n)(x_n+x_{n-1})=x_n^2+2x_n/A-1+2/A,\eqdef\utri$$
the deautonomisation of which was presented in [\dasym], equation 45. Similarly, starting from (\ptes) we introduce
$$Y_n={y_n+1\over Dy_n}\eqdef\utes$$
and find the invariant
$$K={y_ny_{n-1}(y_ny_{n-1}(D^2-1)-y_n-y_{n-1}-1)\over y_n+y_{n-1}},\eqdef\udyo$$
with $\mu=0$, and with the constraint on the parameters $D(C+1/C)=2$. From (\udyo) we obtain the mapping
$$(y_{n+1}+y_n)(y_n+y_{n-1})={y_n^3(D^2-1)+y_n\over y_n(D^2-1)-1}.\eqdef\upen$$
The deautonomisations of (\utri) and (\upen) are, of course, precisely those we find in the $\gamma=0$ case, the important result being that here they are obtained directly for a non-QRT mapping.

A restoration to E$_7^{(1)}$-associated equations is also possible. We start from the remnant equation (\tena) and its invariant (\tdyo), introduce the homographic transformation
$$X_n=\left({d^2-1\over d^2-q}\right){qx_n-d\over x_n-d}\eqdef\fcena$$
and find 
$$K={(qx_n-d)( x_n-d)(qx_{n-1}-d)( x_{n-1}-d)\over(x_nx_{n-1}-q)(x_nx_{n-1}-1)},\eqdef\fcdyo$$
with $\mu=qd^4(d^2-1)(q-1)^4(d^2-q^2)^{-1}(d^2-q)^{-4}$. The initial parameters $A,B$ are expressed in terms of $d,q$ as $A=1-d^2(q-1)^2(d^2-q)^{-2}$ and $B=1-d^2(q-1)^2(q+1)(d^2-q^2)^{-1}(d^2-q)^{-1}$. From (\fcdyo) we obtain the mapping
$$\left({x_{n+1}x_n-q\over x_{n+1}x_n-1}\right)\left({x_{n-1}x_n-q\over x_{n-1}x_n-1}\right)={x_n-q^2/d\over x_n-1/d},\eqdef\fctri$$
which can be deautonomised to
$$\left({x_{n+1}x_n-q_n\over x_{n+1}x_n-1}\right)\left({x_{n-1}x_n-q_{n-1}\over x_{n-1}x_n-1}\right)={x_n-q_nq_{n-1}/d_n\over x_n-1/d_n},\eqdef\stri$$
an equation first derived in [\sequel]. The precise $n$-dependence is given by $\log d_n=3(\alpha n+\beta)-\delta n(-1)^n+\chi_4(n)+\log a$ and $\log q_n=4(\alpha n+\beta)-2\alpha-2\delta (-1)^n+\phi_3(n)$, where $a$ is a constant. We should point out here that the term $\delta n(-1)^n$ is erroneously given in [\sequel]) as $(-1)^n$ rather than  $n(-1)^n$.

Similarly, for the $y$ equation we start from the remnant equation (\ptri) and its invariant (\ptes). We introduce the homographic transformation
$$Y_n={1\over D(z^2+1)}{y_n(D^2z^2+1)-z(D^2+z^2)\over y_n-z}\eqdef\fctes$$
and find
$$K={(y_n-z)(y_{n-1}-z)\big(y_ny_{n-1}(D^2z^4-1)-(y_n+y_{n-1})z^2(D^2-1)+z^2(D^2-z^4)\big)\over (y_ny_{n-1}-1)(y_ny_{n-1}-z^2)}\eqdef\fcpen$$
with $\mu=-Dz^2(D^2-1)^2/(z^2+1)$ and $C+1/C=D+1/D-z^2(D^2-1)^2/(D(z^2+1)^2)$. From the invariant (\fcpen) we obtain the mapping
$$\left({y_{n+1}y_n-z^2\over y_{n+1}y_n-1}\right)\left({y_{n-1}y_n-z^2\over y_{n-1}y_n-1}\right)={(y_n-z)(y_n-z^3)\over y_n^2-Gy_n+1}\eqdef\fchex$$
where $G=\big(D^2(z+1/z)-z^3-1/z^3\big)/(D^2-1)$. The latter can be deautonomised to 
$$\left({y_{n+1}y_n-z_{n+1}z_n\over y_{n+1}y_n-1}\right)\left({y_{n-1}y_n-z_{n-1}z_n\over y_{n-1}y_n-1}\right)={(y_n-z_n)(y_n-z_{n+1}z_nz_{n-1})\over (y_n-a^2)(y_n-1/a^2)},\eqdef\stes$$
an equation obtained here for the first time.
The parameters of equations (\stri) and (\stes) are intimately related and this is essential for the existence of a Miura relation.  We have for instance $q_n=z_{n+1}z_{n-1}$ and $d_nd_{n-1}=a^2z_{n+1}z_nz_{n-1}$. We find thus $\log z_n=2(\alpha n+\beta)-\alpha+\delta (-1)^n-\phi_3(n)+\chi_4(n)+\chi_4(n-1)$. 
Before proceeding further it is convenient to introduce an auxiliary variable $\zeta_n$ related to $z_n, d_n$ by $z_n=\zeta_n\zeta_{n-1}$, $d_n=a\zeta_{n+1}\zeta_n\zeta_{n-1}$. The variable $\zeta_n$ obeys the equation $\zeta_{n+8}\zeta_{n+7}\zeta_{n+1}\zeta_{n}=\zeta_{n+5}\zeta_{n+4}^2\zeta_{n+3}$, the solution of which is $\log \zeta_n=\alpha n+\beta+(\delta  n+\eta)(-1)^n+\chi_4(n)+\phi_3(n-1)$, but $\eta$ can be put to zero through the appropriate gauge (and here the term $n(-1)^n$ is unavoidable). The Miura can now be expressed, in terms of the $\zeta_n$, in the form
 $$x_n=a \zeta_{n} 
{y_{n-1} y_{n} ( - \zeta_{n-2} \zeta_{n-1} + a^2) + y_{n-1} (\zeta_{n-2} \zeta_{n-1}^2 \zeta_{n} - 1) + \zeta_{n-2} \zeta_{n-1} ( - \zeta_{n-1} \zeta_{n} a^2 + 1)\over y_{n-1} y_{n} (\zeta_{n-1} \zeta_{n} a^2 - 1) + y_{n} a^2 ( - \zeta_{n-2} \zeta_{n-1}^2 \zeta_{n} + 1) + \zeta_{n-1} \zeta_{n} (\zeta_{n-2} \zeta_{n-1} - a^2)},\eqdaf\spen$$
$$y_n=\zeta_{n} 
{x_{n+1} x_{n} \zeta_{n-1} \zeta_{n+1} \zeta_{n} ( - \zeta_{n-2} \zeta_{n-1} + a^2) + x_{n+1} a (\zeta_{n-2} \zeta_{n-1} \zeta_{n+1} \zeta_{n} - 1) + \zeta_{n-2} \zeta_{n-1}^2 \zeta_{n+1} \zeta_{n} ( - \zeta_{n+1} \zeta_{n} a^2 + 1)\over x_{n+1} x_{n} (\zeta_{n+1} \zeta_{n} a^2 - 1) + x_{n} \zeta_{n-1} \zeta_{n+1} \zeta_{n} a ( - \zeta_{n-2} \zeta_{n-1} \zeta_{n+1} \zeta_{n} + 1) + \zeta_{n+1} \zeta_{n} (\zeta_{n-2} \zeta_{n-1} - a^2)}.\eqno(\spen b)$$
As expected, multistep evolution equations can also be derived in this case. We find, for instance, for the double-step evolution in $x$ the equation
$$\left({x_{n+2}-q_{n+1}x_n\over x_{n+2}q_{n}-x_n}\right)\left({x_{n-2}-q_{n-2}x_n\over x_{n-2}q_{n-1}-x_n}\right)
={(x_nq_{n-2}-d_{n-1})(x_nq_{n+1}-d_{n+1})(x_n-1/d_n)\over (x_n-d_{n-1})(x_n-d_{n+1})(x_n-q_{n}q_{n-1}/d_n)}\eqdef\shex$$
where $q$ and $d$ are the same as for equation (\stri). This equation was first derived in [\refdef\period], equation (23).

Additive equations related to E$_7^{(1)}$ can also be obtained. Starting from (\tdyo) we introduce the transformation
$$X_n={a(a-1)x_n+a\beta(2-a)\over (a-1)x_n+a\beta},\eqdef\shep$$
where the parameter $a$ is related to both $A$ and $B$ through $A=2a-a^2$ and $B=a(2a-3)/(a-2)$, hence the constraint on the latter: $4A^2+AB^2-6AB-3A+4B=0$. The value of $\mu$ is now $\mu=a(a-1)^4/(a-2)$ and the invariant becomes
$$K={(a(a-1)x_n+a\beta(2-a))((a-1)x_n+a\beta)(a(a-1)x_{n-1}+a\beta(2-a))((a-1)x_{n-1}+a\beta)\over(x_n+x_{n-1})(x_n+x_{n-1}+2\beta)}.\eqdef\soct$$
From (\soct) we obtain the mapping
$$\left({x_{n+1}+x_n+2\beta\over x_{n+1}+x_n}\right)\left({x_{n-1}+x_n+2\beta\over x_{n-1}+x_n}\right)={x_n+\beta(3a-4)/(a-1)\over x_n-a\beta/(a-1)},\eqdef\uenn$$
the deautonomisation of which was presented in [\sequel], equation (3.24).
Similarly we can start from (\ptes) and introducing 
$$Y_n={1\over2D}{(D^2+1)y_n+\beta(3-D^2)\over y_n+\beta},\eqdef\udek$$
obtain the invariant
$$K={(y_n+\beta)(y_{n-1}+\beta)((D^2+1)y_n+\beta(3-D^2))(y_n+\beta)((D^2+1)y_{n-1}+\beta(3-D^2))\over (y_n+y_{n-1})(y_n+y_{n-1}+2\beta)},\eqdef\wena$$
with $\mu=-D(D^2-1)/4$ and the parameter constraint $4D(C+1/C)+D^4-6D^2-3=0$ (which is just the transcription of the constraint between $A$ and  $B$ obtained above given the the relations between $A, B$ and $C,D$). The corresponding mapping is
$$\left({y_{n+1}+y_n+2\beta\over y_{n+1}+y_n}\right)\left({y_{n-1}+y_n+2\beta\over y_{n-1}+y_n}\right)={(y_n+\beta)(y_n+3\beta)\over y_n^2-\beta^2(D^2-9)/(D^2-1)},\eqdef\wdyo$$
the deautonomisation of which follows closely that of (\stes).

\bigskip
5. {\scap Another interesting limit}

\medskip
In the previous sections we established the relation between the equation IX of [\multi] and the symmetric equations also obtained in the same paper, cases II and V. In this section we will show that case IX is a kind of master one, since it allows to obtain case XII of [\multi] as a special limit, to which we shall apply our restoration procedure. The case XII in question is a system of the form (\ddyo) with $z_n=\eta_{n}-\gamma+\phi_3(n-1)$, $\zeta_n=\eta_{n}+\gamma+\phi_3(n+1)$, $k_n=\chi_4(n)$, $\kappa_n=\delta+\phi_2(n)$ with $\eta_n=\alpha n+\beta$. 

We start from case IX and neglect the  periodic terms with the exception of the $\phi_2(n)$ one. For the latter we assume that it has exactly the amplitude $\gamma$, i.e. $\phi_2(n)=\gamma(-1)^n$, which means that for even indices $\kappa_n$ is equal to $2\gamma$ and for odd ones equal to 0. 
Neglecting the secular dependence and scaling $\beta$ to 1 we obtain
$${x_{n+1}-(4-\phi_2(n)-\gamma)^2\over x_{n+1}-(2-\phi_2(n)+ \gamma)^2}{x_{n}-(4-\phi_2(n)+ \gamma)^2\over x_{n}-(2-\phi_2(n)- \gamma)^2}{y_{n}-(1-\phi_2(n))^2\over y_{n}-(5-\phi_2(n))^2}=1,\eqdaf\wone$$
$${y_{n}-(1+\phi_2(n)+2 \gamma)^2\over y_{n}-(1-\phi_2(n))^2}{y_{n-1}-(1+\phi_2(n))^2\over y_{n-1}-(1-\phi_2(n)-2 \gamma)^2}{x_{n}-(2-\phi_2(n)- \gamma)^2\over x_{n}-(2+\phi_2(n)+ \gamma)^2}=1.\eqno(\wone b)$$
Next we eliminate the variable $y$ and obtain a single equation for $x$ which can be cast into trihomographic form as
$${x_{n+1}-(2-\phi_2(n)+ \gamma)^2\over x_{n+1}(2+\phi_2(n)- \gamma)^2}{x_{n-1}-(2-\phi_2(n)+ \gamma)^2\over x_{n-1}-(2+\phi_2(n)- \gamma)^2}{x_{n}-(2+\phi_2(n)-\gamma)^2\over x_{n}-(4-\phi_2(n)+\gamma)^2}=1,\eqdef\wdyo$$
Separating this equation for even and odd indices $n$ we remark that due to the presence of the quantity $\phi_2(n)-\gamma$ the equation collapses for even $n$. Computing carefully the limit we find that it reduces to 
$${1\over x_{2n+1}-1}+{1\over x_{2n-1}-1}-{2\over x_{2n}-4}=0.\eqdaf\wtri$$
On the other hand, the equation for odd $n$ can be computed without any problem leading to 
$${x_{2n}-(1+ \gamma)^2\over x_{2n}-(1- \gamma)^2}{x_{2n-2}-(1+ \gamma)^2\over x_{2n-2}-(1- \gamma)^2}{x_{2n-1}-(2-\gamma)^2\over x_{2n-1}-(2+\gamma)^2}=1.\eqno(\wtri b)$$
In both equations above we have multiplied the variable $x$ by 4.
Computing now the autonomous limit of case XII we find 
$${1\over x_{n+1}-1}+{1\over x_{n}-1}-{2\over y_n-4}=0,\eqdaf\wtes$$
$${y_n-(1+\delta)^2\over y_n-(1-\delta)^2}{y_{n-1}-(1+\delta)^2\over y_{n-1}-(1-\delta)^2}{x_{n}-(2-\delta)^2\over x_{n}-(2+\delta)^2}=1.\eqno(\wtes b)$$
The parallel between (\wtri) and (\wtes) is clear: $x$ and $y$ of (\wtes) correspond to the odd and even index $x$ of (\wtri) and $\gamma$ in the latter is the $\delta$ of the former. 
(While equation (\wtes) was in principle derived in [\multi] an unfortunate misprint gave a wrong form for (\wtes a)).
In order to obtain the remnant equation we start from (\wtes) and introduce a homographic transformation for both variables in order to simplify it to the maximum: 
$$X_n={(1+\delta)(3+\delta)\over 8\delta}{x_n-(2-\delta)^2\over x_n-1},\qquad Y_n=c{y_n-(1+\delta)^2\over y_n-(1-\delta)^2},\eqdef\fec$$
 where $c^2=(1-\delta)(3-\delta)/((1+\delta)(3+\delta))$. The final equation is
$$X_n+X_{n+1}=B+{1\over 1-AY_n},\eqdaf\wpen$$
$$Y_nY_{n-1}=1-{1\over X_n}\eqno(\wpen b)$$
where $X$ and $Y$ are homographic in terms of $x$ and $y$ and $A,B$ are expressed in terms of $\delta$ as
$A=(1+\delta)(3-\delta)/(c(1-\delta)(3+\delta))$ and $B=(3+\delta)/(2\delta)$.
They are related by the constraint
$$A^2B(B-2)^3-(B-1)(B+1)^3=0.\eqdef\whex$$
As is customary, we shall in what follows, consider  $A$ and $B$ to be two independent parameters without the constraint (\whex).

Equation (\wpen) is a QRT mapping and possesses an invariant in the form
$$K={X_n^2Y_n^2-BX_nY_n^2-X_n^2Y_n(A+1/A)+X_nY_n(AB+(B+1)/A)+X_n^2-X_n(B+1)+B\over Y_n}.\eqdef\whep$$
Moreover, deautonomising (\wpen) is straightforward. We find
$$X_n+X_{n+1}=B_n+{Z_{n+1}+Z_{n-1}\over 1-AY_n},\eqdaf\woct$$
$$Y_nY_{n-1}=1-{Z_{n}+Z_{n-1}\over X_n}\eqno(\woct b)$$
where $Z_n=\alpha n+\beta+\phi_3(n)$, $B_n=Z_n+b$, while $A, b$ are constant, leading to an equation associated to the affine Weyl group D$_4^{(1)}$.  

Given the form of (\wpen) we can easily write the equations for the variables $X$ and $Y$ alone. For the former we start by translating $X\to X+B/2$ leading to the canonical form
$$\left({X_n+X_{n+1}-1\over X_n+X_{n+1}}\right)\left({X_n+X_{n-1}-1\over X_n+X_{n-1}}\right)=A^2\left(1-{1\over X_n+B/2}\right).\eqdef\wenn$$
Its deautonomised form is
$$\left({X_n+X_{n+1}-Z_{n+1}-Z_{n-1}\over X_n+X_{n+1}}\right)\left({X_n+X_{n-1}-Z_{n}-Z_{n-2}\over X_n+X_{n-1}}\right)=A^2\left({X_n-\rho_{n+1}-\rho_n-\rho_{n-1}+2b\over X_n+\rho_n}\right),\eqdef\wdek$$
where the $Z_n$, $B_n$ and $A$ are the same as for (\woct) and $\rho$ is an auxiliary quantity given by $\rho_n=\alpha n/2+\delta-\phi_3(n+1)$ with $\delta=b/2+\beta/2-\alpha/4$ and thus $B_n= \rho_n+\rho_{n+1}$.

The equation for $Y$ can be obtained in canonical form directly from (\wpen):
$${1\over 1-Y_nY_{n+1}}+{1\over 1-Y_nY_{n+1}}=B+{1\over 1-AY_n},\eqdef\oena$$
and in non autonomous form
$${Z_{n}+Z_{n+1}\over 1-Y_nY_{n+1}}+{Z_{n}+Z_{n-1}\over 1-Y_nY_{n+1}}=B_n+{Z_{n+1}+Z_{n-1}\over 1-AY_n},\eqdef\odyo$$
with the same $Z_n$, $B_n$ and $A$ as for (\woct) and (\wdek). Both equations (\wdek) and (\odyo) are associated to D$_4^{(1)}$.
No double-step evolution equations are possible here. In fact the equations we derived, given the parallel between (\wtri) and (\wtes), are in some sense already double-step evolutions (and in the same sense a triple-step evolution does not exist here, since it would have been a sextuple-step one).

Starting from the remnant equation (\wpen) we can apply the restoration process and obtain  discrete Painlev\'e equations. Clearly when the parameters $A$ and $B$ obey the relation (\whex) we can go back to the initial additive E$_8^{(1)}$-associated equation, namely the one we call case XII, while removing the constraint (\whex) leads, as expected, to the multiplicative version of case XII. Just as in the $\gamma=0$ case, no restoration towards E$_7^{(1)}$ associated equations exists.
However a restoration towards equations associated with the group E$_6^{(1)}$ is possible. In this case we introduce the homographic transformation
$$X_n={ABx_n+1\over Ax_n+1-A^2}\eqdaf\otri$$
$$Y_n={Ay_n+B-1\over y_n+AB}\eqno(\otri b)$$
and obtain the system 
$$(x_ny_n-1)(x_{n+1}y_n-1)=(1-y_n(A-1/A))^2,\eqdaf\otes$$
$$(x_ny_n-1)(x_ny_{n-1}-1)=(1+x_n(B-1)/A)(1+x_nAB).\eqno(\otes b)$$
However (\otes) is, up to trivial scalings, the same as equation (\vhep) we encountered in Section 3 and which was deautonomised to (\venn). Thus the restoration from (\wpen) to a E$_6^{(1)}$-associated equation does not lead to any new result. Clearly the same holds true for the restoration to an additive E$_6^{(1)}$-related equation.

\bigskip
6. {\scap Conclusions}

\medskip
In this paper we set out to construct discrete Painlev\'e equations, starting from equations that are associated to the affine Weyl group E$_8^{(1)}$. The method we used for this purpose, which we named restoration, comprises two phases. We start with some discrete Painlev\'e equation and first take its autonomous limit to a QRT mapping. (Any discrete Painlev\'e equation could be considered at this stage but starting from an equation associated to E$_8^{(1)}$ is convenient since all the E$_8^{(1)}$ related deautonomisations  are then known quasi automatically and one does not have to derive these complicated systems from scratch). Once the autonomous mapping is obtained we introduce a homographic transformation in order to bring it to its simplest possible form, which we call the remnant mapping. This concludes the first phase,  the so-called deconstruction phase. Starting from the invariant of the remnant mapping we then introduce homographic transformations and try to obtain all possible invariants in one of the canonical forms we classified in [\canon]. From these invariants we obtain mappings which  are then to be deautonomised to discrete Painlev\'e equations. This is the so-called restoration phase.

In this paper we have worked with the discrete Painlev\'e equation, already derived in [\deight], given by equation (\ddyo). The coefficients appearing in this equation have periods 2, 3 and 4. When after restoring a particular form of the remnant equation, the resulting discrete Painlev\'e equation has fewer periods appearing in its coefficients than were present in the original equation (\ddyo), one must conclude that the equation has lost the corresponding (genuine) degrees of freedom that were present in the original, (where by `genuine degrees of freedom' we refer to the parameters the values of which can be modified through Schlesinger transformations). For instance, the disappearance of the period 2 leads to a loss of one degree of freedom and an equation associated with E$_7^{(1)}$. Similarly when the period 3, respectively 4, disappears we find equations related to E$_6^{(1)}$, respectively D$_5^{(1)}$.

As we have seen here, the simultaneous loss of the period 4, the period 2 and the free parameter $\gamma$ leads to an equation associated with A$_3^{(1)}$. However, as we have shown in section 5, by choosing the value of the free parameter $\gamma$ in a specific relation to the amplitude of the period-two function in the coefficients, it is possible to lose only one degree of freedom in addition to the period 4, leading to a D$_4^{(1)}$ associated equation. In the case of E$_8^{(1)}$, E$_7^{(1)}$ or E$_6^{(1)}$, one can obtain either an additive or a $q$-difference equation (and an elliptic one for the former). Thus the ``higher" part of the diagram in the introduction is well represented in this analysis.

The choice of the system (\ddyo) was not innocuous. While the autonomous limit of the equation for $\gamma=0$ indeed leads to a QRT mapping, the general case with $\gamma\neq0$ gives rise to a non-QRT mapping. Thus the restoration method cannot be applied, as such. However, as we have shown in section 4, it is possible to circumvent this difficulty. It suffices to eliminate either of the two variables of the mapping in order to obtain, in fine, a QRT mapping for which the restoration procedure can be applied. Once the various discrete Painlev\'e equations were obtained we showed that for each of them it is possible to construct a Miura transformation linking the equations for the two different variables. This Miura system is of course a discrete Painlev\'e equation in its own right (but possibly in non-canonical form) and is in fact the equation which we would have expected to obtain from the restoration process. We believe that this is not a feature limited to the equation at hand but rather one that would characterise all E$_8^{(1)}$-associated equations which possess constant parameters.

Another interesting result of this paper is the construction of multistep evolution equations starting from a single-step one. The possibility for a double-step or triple-step evolution depends crucially on the system at hand. Seeking multiple step evolutions is another mechanism for  constructing discrete Painlev\'e equations, not directly related to the restoration method, but one which can also lead to most interesting results and in particular new discrete Painlev\'e equations. We intend, in future works of ours, to pursue this line of research which till now has proven particularly fruitful.

\bigskip
{\scap Acknowledgements}

\medskip
RW would like to acknowledge support from the Japan Society for the Promotion of Science (JSPS) through JSPS grant number 18K03355.

\bigskip
{\scap References}
\medskip

[\dps] A. Ramani, B. Grammaticos and J. Hietarinta, Phys. Rev. Lett. 67 (1991) 1829.

[\infin] A. Ramani and B. Grammaticos, Phys. Lett. A 373 (2009) 3028.

[\brezov] E. Br\'ezin and V.A. Kazakov, Phys. Lett. 236B (1990) 144.

[\shohat] J.A. Shohat, Duke Math. J. 5 (1939) 401.

[\pondy] K.M. Tamizhmani, T. Tamizhmani, B. Grammaticos and A. Ramani, Springer LNP 644 (2004) 323.

[\sincon] B. Grammaticos, A. Ramani, V. Papageorgiou, Phys. Rev. Lett. 67 (1991) 1825.

[\capel] A. Ramani and B. Grammaticos, Physica A 228 (1996) 160.

[\sakai] H. Sakai, Commun. Math. Phys. 220 (2001) 165.

[\take] T. Takenawa, J. Phys. A: Math. Gen. 34 (2001) 10533.

[\mase] T. Mase, J. Integr. Sys. 3 (2018) xyy010.

[\kundu] K.M. Tamizhmani, B. Grammaticos, A. Ramani and T. Tamizhmani, in Classical and Quantum Integrable Systems (IOP Publishing) ed. A. 
Kundu. (2003)  p64.

[\rains] E.M. Rains, {\sl Generalized Hitching systems on rational surfaces}, preprint (2016)  arXiv:1307.4033v2.

[\resto] R. Willox, A. Ramani and B. Grammaticos, J. Math. Phys. 58 (2017) 123504.

[\qrt] G.R.W. Quispel, J.A.G. Roberts and C.J. Thompson, Physica D34 (1989) 183.

[\deight] A. Ramani and B. Grammaticos, J. Phys. A 48 (2015) 355204.

[\auton] A. Ramani, S. Carstea, B. Grammaticos and Y. Ohta, Physica A  305 (2002) 437.

[\canon] A. Ramani, B. Grammaticos, J. Satsuma and T. Tamizhmani, J. Phys. A 51 (2018) 395203.

[\multi] B. Grammaticos and A. Ramani, J. Phys. A 48 (2015) 16FT02.

[\dress] B. Grammaticos, A. Ramani and V. Papageorgiou, Phys. Lett. A 235 (1997) 475.

[\cfive] T. Tokihiro, B. Grammaticos and A. Ramani, J. Phys. A 35 (2002) 5943.

[\japan] B. Grammaticos, T. Tamizhmani, A. Ramani, A. S. Carstea  and K.M.Tamizhmani, J. Phys. Soc. Japan 71 (2002) 443.

[\congr] T. Tamizhmani, B. Grammaticos, A. Ramani and K.M .Tamizhmani, Physica A 369 (2006) 463.

[\ohta] B. Grammaticos, Y. Ohta, A. Ramani and H. Sakai, J. Phys. A 31 (1998) 3545.

[\addit] A. Ramani and B. Grammaticos, J. Phys. A 50 (2017) 055204.

[\qasym] B. Grammaticos, A. Ramani, K.M. Tamizhmani, T. Tamizhmani and J. Satsuma, J. Math. Phys. 57 (2016) 043506.

[\dasym] B. Grammaticos, A. Ramani, K.M. Tamizhmani, T. Tamizhmani and J. Satsuma, J. Math. Phys. 55 (2014) 053503.

[\sequel] A. Ramani, R. Willox, B. Grammaticos, A.S. Carstea and J. Satsuma, Physica A 347 (2005) 1.

[\period] A. Ramani and B. Grammaticos, J. Phys. A FT 47 (2014) 192001.

\end{document}